%% file: article.tex
\documentclass[preprint, 12pt]{elsarticle}
\usepackage{amsmath}

\usepackage{amssymb}
\usepackage{pifont}
\usepackage{natbib}
\usepackage{geometry}
\usepackage{fleqn}
\usepackage{graphicx}
\usepackage{txfonts}
\usepackage{hyperref}

\begin{document}

\begin{frontmatter}

\title{The fraud loss for selecting the model complexity in fraud detection}

\author{Simon Boge Brant\corref{cor1}}
\ead{simonbb@math.uio.no}
\cortext[cor1]{Corresponding author}
\author{Ingrid Hob\ae k Haff}

\address{}

\begin{abstract}
  \input{abstract.tex}
\end{abstract}

\end{frontmatter}

\section{Introduction}
\noindent\input{introduction.tex}

\section{Models}
\label{sec:mods}
\noindent
\input{models.tex}
\section{Problem description}
\label{sec:prob}
\noindent
\input{problem_description.tex}
\section{Selecting the model complexity}
\label{sec:modsel}
\noindent
\input{sel_mod_comp.tex}
\section{Simulation study}
\label{sec:simstud}
\noindent
\input{simulating_data.tex}\label{subsec:simdata}
\noindent
\input{simulation.tex}
\section{Illustration on VAT fraud data}
\label{sec:realdat}
\noindent
\input{illustration.tex}
\section{Concluding remarks}
\label{sec:concl}
\noindent
\input{conclusion.tex}
\section{Acknowledgements}
\input{acknowledgements.tex}
\newpage
\bibliography{article}
\end{document}

%% file: abstract.tex
  In fraud detection applications, the investigator is typically limited to
  controlling a restricted number $k$ of cases. The most efficient manner of
  allocating the resources is then to try selecting the $k$ cases with the
  highest probability of being fraudulent. The prediction model used for this
  purpose must normally be regularized to avoid overfitting and consequently
  bad prediction performance. A new loss function, denoted the fraud loss, is
  proposed for selecting the model complexity via a tuning parameter. A
  simulation study is performed to find the optimal settings for validation.
  Further, the performance of the proposed procedure is compared to the most
  relevant competing procedure, based on the area under the receiver operating
  characteristic curve (AUC), in a set of simulations, as well as on a VAT fraud
  dataset. In most cases, choosing the complexity of the model according to the
  fraud loss, gave a better than, or comparable performance to the AUC in terms
  of the fraud loss.

%% file: introduction.tex
Fraud detection has cropped up as a term in statistical research at least
since the early 1990s. In the excellent review article by 
\citet{bolton2002statistical}, the authors identify some of the characteristics
that make statistical fraud detection a distinct field of the statistical
literature, and not just a special case of binary classification, or some other
well-understood problem class. The goal of statistical fraud detection, is to
create a system that automatically selects a subset of all cases, (insurance
claims, financial transactions, etc.), that are the most interesting cases for
further investigation. This is necessary because there is typically a much
higher number of claims than one realistically could investigate manually, and
because fraud is typically quite rare. In simple terms, statistical fraud
detection can be thought of as binary classification, potentially with highly
imbalanced classes, depending on the type of fraud. By imbalanced classes we
mean, in this case, that there are very few occurrences of one of the two
possible outcomes. In other words, the vast majority of financial transactions 
or insurance claims are legitimate, and fraud is, relatively speaking, a rare
occurrence. 

In fraud detection applications, the investigator is often required to 
efficiently allocate limited resources. This amounts to selecting a restricted 
number of cases, those that are most likely to be fraudulent, or most worthy
of investigation. In order to achieve this, a model should be fitted to recorded
data of previously investigated cases, and then used to predict the 
probability of fraud on new cases. The set of cases to be investigated should
subsequently be determined from the predicted probabilities. In this respect,
we have a precise notion of what a good, or bad, model is for this purpose, 
namely one that lets us pick a certain number of cases, such that as many as 
possible of these are actual cases of fraud. Given the application, we term 
this notion \textit{fraud loss}. However, we acknowledge that it has been 
studied in other contexts under different names, notably by
\citet{clemenccon2007ranking}, where they refer to the problem as
\textit{finding the best instances}, or \textit{classification with a mass
constraint}. 

The problem of minimising fraud loss, or \textit{finding the best instances} is
equivalent to maximising a measure known as the \textit{precision at k} in the field
of information retrieval. This is discussed amongst others by
\citet{robertson2007rank}, \citet{joachims2005support}, and
\citet{NIPS2012_4635}. A related problem is that of local bipartite ranking,
where the aim is to find the best pairwise ranking of a subset of the data. In
the language of \citet{clemenccon2007ranking}, the focus is then not only on
\textit{finding} the best instances, but to \textit{rank the best instances.}
The goal in this setting is not only to select the $k$ most relevant instances, 
but also to rank them as well as possible. In the context of fraud, this 
corresponds to selecting $k$ cases such that as many as possible are positive, 
and that if they are ordered according to their predicted probability of fraud, 
the greatest possible number of the selected positive cases are ranked higher 
than the selected negative cases. 

There are several suggestions for how to estimate a model in order to solve 
these and related problems. \citet{NIPS2012_4635} propose an estimation 
criterion that should result in model that finds the best instances. This
criterion involves minimising a hinge-type loss function over all pairs of 
observations with opposite outcome. The number of optimisation problems to solve
in the estimation procedure is then twice the number of pairs.
\citet{rudin2009p} proposes an estimation framework that concentrates on the
top ranked cases, called the \textit{P-norm push}. Her method is inspired by the
\textit{RankBoost} algorithm of \citet{freund2003efficient} for minimising the
ranking loss, and is an extension of \textit{RankBoost} to a more general
class of objective functions. \citet{eban2016scalable} propose estimation
methods aiming to maximise different measures relevant to ranking, such as the
area under the precision-recall curve, and the recall at a fixed precision. They
do this by approximating the false positive, and true positive rates in the
objective function. Their aim is to construct the objective function in such a
way that the derivatives have more or less the same complexity as those of the
log-likelihood function of a logistic regression model. This makes the method a
lot more scalable than those of \citet{rudin2009p}, and \citet{NIPS2012_4635}.

As opposed to the papers mentioned above, we will not focus on the estimation
of the model parameters directly, but rather on choosing the complexity of the
model, via a tuning parameter. More specifically, we will consider maximisation
of the likelihood function of the statistical model with regularisation, using
penalised methods, or boosting. Different values of the regularisation
parameters will then result in models of varying complexity. The model is to be
used for a very specific purpose, namely to make predictions in order to select
the $k$ most likely cases of fraud among a new set of cases. Therefore, it seems
reasonable to try to choose the regularisation parameters that are optimal for
this particular application. In that context, we define a loss function, which, 
broadly speaking, is the number of non-fraudulent cases among the $k$ selected.

An important question, is how to estimate the out of sample value of the fraud 
loss function, in order to select tuning parameters. There are a number 
of different validation techniques that one may employ to mimic a new dataset,
using the training data. They all involve fitting models to different subsets 
of the data, and evaluating the error on the data points that are left out. As
the application is somewhat special, standard settings and techniques may not
be adequate. For instance, predicted class labels will depend on an empirical 
quantile of the predicted probabilities, which might require subsamples of a
certain size to be stable. Therefore, we will investigate what the best 
strategies for out of sample validation are.

The paper is organized as follows. Section \ref{sec:mods} defines the models
that form the basis of the problem, whereas Section \ref{sec:prob} describes the
actual problem. Further, the approach for selecting the model complexity with 
fraud loss is presented in Section \ref{sec:modsel}. In Section
\ref{sec:simstud}, the properties of the proposed method is evaluated in a
simulation study, and the method is further tested on real data in Section
\ref{sec:realdat}. Finally, Section \ref{sec:concl} provides some concluding
remarks.

%% file: models.tex
We want our models to produce predicted probabilities of fraud, or at least
predicted scores, with the same ordering as the probabilities. In order to
estimate probabilities of fraud, one should find a model that maximises the
likelihood of the data. In what follows, $Y$ denotes the binary outcome, which
is an indicator of whether a specific case is fraudulent, and $\mathbf{X}$ 
represents the $p$-dimensional random vector of covariates. We will in all cases consider
models where
\begin{align*}
  \log\left(\frac{\text{Pr}(\mathbf{Y} = 1 \vert \mathbf{X}=\mathbf{x})}
    {\text{Pr}(Y=0\vert \mathbf{X}=\mathbf{x})}\right) = 
  \eta(\mathbf{x}).
\end{align*}
The model $\eta(\mathbf{x})$ could be a linear function of the covariates,
or an additive model where each component is a regression tree. This
specification implies that the conditional probabilities of an event will take the 
form
$$ 
p_i = \text{Pr}\left(Y_i=1\vert \mathbf{X}_i=\mathbf{x}_i\right) =
\frac{\exp(\eta_i)}{1 + \exp(\eta_i)},
$$ 
where $\hat\eta_i$ is a shorthand for $\hat\eta(\mathbf{x}_i).$  The fraud 
indicators $Y_{i}$ are assumed to be conditionally independent, and 
Bernoulli-distributed, with probabilities $p_i$, given $\mathbf{X}_i=\mathbf{x}_i$.
That leads to a binary regression model with a logit link
function, that has the associated log-likelihood function
\begin{align*}
 ll &= \sum_{i=1}^n\log\left({p}_i^{y_i}(1 - {p}_i)^{1 - y_i}\right) = \sum_{i=1}^n \left(y_i\log\left(\frac{{p}_i}{1 - {p}_i}\right)- \log(1 - {p}_i)\right) = \sum_{i=1}^n\left(y_i\eta_i - \log(1 - {p}_i)\right).
\end{align*}
In many cases, for instance if the covariates are noisy, or high-dimensional, 
the predictive accuracy of the model can be improved by shrinking the
predicted probabilities towards the common value $\frac{1}{n}\sum_{i=1}^{n}y_i$,
which is the estimate of the marginal probability $\text{Pr}(Y=1)$. In the case of a
parametric linear model, this would correspond to shrinking the regression
coefficients towards zero, and for a tree model to a less complex model, in
terms of the number of trees, the depth of each tree, and possibly also the
weights assigned to the leaf nodes. One option, in the case of the linear model
$\eta(\mathbf{x}) = \beta_{0}+\boldsymbol{\beta}^{t}\mathbf{x}$,
is to only look for solutions where  $(\hat \beta_{0}, \boldsymbol{\hat\beta})$ is at 
most some distance from the origin, as measured by a euclidean distance. This 
regularised estimator is the maximiser of the penalised log-likelihood function
\begin{equation}
ll(\beta_{0},\boldsymbol{\beta})-\lambda\sum_{j=1}^p\hat \beta_j^2 = \sum_{i=1}^n\left(y_i\eta_i - \log(1 - {p}_i)\right)-\lambda\sum_{j=1}^p\hat \beta_j^2,
\label{loglik_lin}
\end{equation}
and is the solution to  
\begin{align*}
 &\underset{\hat \beta_0,\; \boldsymbol{\hat \beta}}{\text{argmax}}
 \sum_{i=1}^{n} y_i\hat\eta_i- \log(1 - \hat p_i) - \lambda\sum_{j=1}^p\hat \beta_j^2.
\end{align*}
It is often referred to as ridge regression \citep{hoerl1970ridge}, or $L_2$-penalised
logistic regression, since it assigns a penalty to the squared $L_2$-norm of the
parameters. Similar estimators such as the lasso
\citep{tibshirani1996regression}, or the elastic net
\citep{zou2005regularization}, can be constructed by considering different norms,
$L_1$ in the case of the lasso, and a convex combination of $L_1,$ and squared
$L_2$ norm, in the case of the elastic net.

We will also consider additive models, that is models where
\begin{equation}
\eta(\mathbf{x}) = f_0 + \sum_{j=1}^M f_j(\mathbf{x}).
\label{tree_model}
\end{equation}
Here, $f_0$ is a constant function, and each $f_j(\mathbf{x})$ is a regression
tree. I.e., a function that for a partition $\{R_t\}_{t=1}^T$ of $\mathbb{R}^p,$ 
takes a constant value $c_t$ for all $\mathbf{x}$ in each $R_t,$ such that
$$f_j(\mathbf{x}) = \sum_{t=1}^Tc_t\mathcal{I}(\mathbf{x}\in R_t).$$

These models are typically fit by gradient boosting \citep{friedman2001greedy},
an iterative procedure where one starts with a constant, then adds one component
at a time, by maximising a local Taylor approximation of the likelihood around
the current model. Usually, this update is scaled down by a factor, i.e.,
multiplied by some number $\nu \in (0, 1),$ in order to avoid stepping too far
and move past an optimum of the likelihood function. Several highly efficient
implementations to fit such models exist, such as LightGBM
\citep{ke2017lightgbm}, CatBoost \citep{dorogush2018catboost}, and XGBoost
\citep{chen2016xgboost}. The flexibility that such models offer, especially if
the individual trees are allowed to be complex, will easily lead to models that
capture too much of the variability in the training data. In order to avoid
this, and get a stable model, we will control the total number of components in
the model, and constrain the complexity of each tree.

%% file: problem_description.tex
An informal description of the fraud detection problem was given in the 
introduction. Here, the problem will be defined and explained in more detail.
Formally, we can describe our version of a fraud detection problem as follows.
We have two datasets, a training set
$$\mathcal{D}^{tr} = \{(Y_i, \mathbf{X}_i)\}_{i=1}^{n_{tr}},$$
that consists of previously investigated cases, and a test set
$$\mathcal{D}^{te} = \{(Y_i, \mathbf{X}_i)\}_{i=1}^{n_{te}},$$
that consists of cases that are yet to be investigated. The $(p+1)$ dimensional 
random vectors $(Y_i, \mathbf{X}_i), \, i=1, ..., n_{tr} + n_{te}$ are assumed to
be iid, and the main interest is the conditional distribution of $Y_i,$ given 
$\mathbf{X}_i.$ In what follows, $n$ will denote for the size of a sample, regardless 
of whether the sample in question is the test set or the training set, unless  it is
unclear from the notation which one is referred to. In some cases, if the data 
contains detailed information of past cases, it could be useful to describe $Y_i$ 
as a categorical or an ordinal random variable. However, we will concentrate on the
binary case, so that each $Y_i$ takes either the value $0$ or $1$. Since the goal
of the investigation is to uncover fraud, there should be as many actual cases of 
fraud as possible among the ones selected for investigation. This
amounts to producing $n$ predictions $\{\widehat Y_i\}_{i=1}^{n},$ such that $k$
of these have the value $1$. Therefore, the minimiser of the loss function is a model 
that minimises
\begin{align*}
  \mathcal{L}^{fraud} = \sum_{i=1}^{n}(1 -  Y_i)
  \widehat Y_i, \text{ s.t. } \sum_{i=1}^{n}\widehat Y_i = k.
\end{align*}
This is equivalent to minimising the classification error 
\begin{align*}
\mathcal{L}^{class} = \sum_{i=1}^{n}\vert Y_i - \widehat Y_i\vert =
\sum_{i=1}^{n}(1 -  Y_i)\widehat Y_i + 
\sum_{i=1}^{n}Y_i(1 - \widehat Y_i),
\end{align*}
under the same constraint. Since $\sum_{i=1}^{n}\widehat{Y_i} = k,$
\begin{align*}
\mathcal{L}^{fraud} = \sum_{i=1}^{n}(1 -  Y_i)\widehat Y_i = k - \sum_{i=1}^{n}Y_i
\widehat{Y_i},
\end{align*}
and
\begin{align*}
\mathcal{L}^{class} &= \sum_{i=1}^{n}\vert Y_i - \widehat Y_i\vert=k + \sum_{i=1}^{n}Y_i - 2\sum_{i=1}^{n}Y_i\widehat{Y_i},
\end{align*}
so the two must have the same minimiser.

The idea is to minimise the expected value of $\mathcal{L}^{fraud}_{te}$, which is 
the fraud loss for the test set. It would therefore be illuminating to know what the
minimiser is, i.e., what it is that one is attempting to estimate. We can write
\begin{align*}
  \mathbb{E}\left(\mathcal{L}_{te}^{fraud}\right) &=
  \mathbb{E}\left(\sum_{i=1}^n(1 - Y_i)\widehat{Y}_i\right)= \mathbb{E}\left(\mathbb{E}
  \left(\sum_{i=1}^n(1 - Y_i)\widehat{Y}_i\Big\vert\mathbf{X}_{te}\right)\right)= \mathbb{E}
  \left(\sum_{i=1}^n(1 - P_i)\widehat{Y}_i\right),
\end{align*}
where $P_i = Pr(Y_i=1\vert \mathbf{X_i}).$ The minimiser of the above expectation 
over all vectors $\widehat{\mathbf{Y}},$ having all elements equal to zero, except 
exactly $k$ elements that are equal to one, is the vector $\widehat{\mathbf{Y}}^*$ 
satisfying
$$\widehat{Y}^*_i = \mathcal{I}(P_i \geq P_{(n - k + 1)}),$$
where $P_{(1)} \leq\ldots\leq P_{(n)}$ are the conditional fraud probabilities $P_{i}$
for the test set, sorted in ascending order. This is an indicator of whether $P_{i}$ 
is among the $k$ largest in the test sample. Thus, a quite natural approach is to 
fit a model for the regression function
$$p(\mathbf{x})= \text{Pr}(Y=1\vert \mathbf{X}=\mathbf{x}),$$
resulting in the estimated probabilities $\widehat{p}_i$, and then use the prediction
$$\widehat{Y}_i = \mathcal{I}(\widehat{p}_i \geq \widehat{p}_{(n - k + 1)}).$$

%% file: sel_mod_comp.tex
We want to find the model that minimises the fraud loss function. This is done by
fitting a sequence of models, either by maximising \eqref{loglik_lin} for a sequence
of values of the penalty parameter $\lambda$, or by fitting the additive tree  model 
\eqref{tree_model} via gradient boosting, where each model has a
different number $M$ of components. Choosing the best of these corresponds to
selecting a value of $\lambda$, in the former case, or $M,$ in the latter case. 
However, the main interest is not the best possible fit to the training data, but 
rather the best possible performance on a new dataset. This may be determined
by estimating the relative out of sample performance of each model, via repeated
cross validation, or bootstrap validation. As mentioned in the introduction, we
will study how different validation schemes perform. Both bootstrap validation 
and cross validation involve splitting the data in different subsets, and repeatedly
using some of the data for fitting, and the rest to evaluate the fitted models.
Since the model is evaluated on datasets whose size $n_{eval}$ is not, in general, the
same as the one $n_{te}$ of the test set, we let the number of selected observations
$k$ be the nearest integer to $\tau n_{eval},$ where $\tau=\frac{k}{n_{te}}\in (0, 1),$ is
the proportion of the cases in the test set we want to select. 

We evaluate the classification error using cross validation with $L$ folds, and
$D$ repetitions. The data are split into different folds by randomly assigning  
observations to each fold, thus creating $D$ sequences of $L$
non-overlapping subsets of the integers $\{1, 2, \dots, n\},$ which we denote as
$\{A_l^{(d)}\}_{l=1}^{L}, d=1, \dots, D.$ The cross-validation statistic is given by
\begin{align*}
  \widehat{\mathcal{L}}^{fraud}_{CV} =
  \frac{1}{D}\sum_{d=1}^D\frac{1}{L}\sum_{l=1}^{L}
  \frac{1}{\vert A_l^{(d)} \vert}\frac{\sum_{i \in A_l^{(d)}}(1 - y_i)\hat y_i^{l, d}}
  {\sum_{i \in A_l^{(d)}} \hat y_i^{l, d}}.
\end{align*}

An alternative to cross validation is bootstrap validation, where one for each
fold draws observations from the training set with replacement, usually as many 
as the number of training observations. The left-out observations from each fold
are then used for validation. The probability that a specific observation is
left out of a bootstrap fold, when the bootstrap folds are of the same size as
the training set, is equal to $\left(1 - \frac{1}{n}\right)^{n},$ which when
$n$ increases, converges to $e^{-1}\approx 0.368.$ This means that the
validation sets on average will contain a little over a third of the total
training data. 

In a standard binary classification setting, one can compute a
statistic that mimics a leave one out cross validation error as
\begin{align*}
  \widehat{\text{Err}^{(1)}} = \frac{1}{n}\sum_{i=1}^n
  \frac{\sum_{b=1}^B \text{I}_{i}^b\text{L}(y_i, \widehat{y}_i^{*b})}
  {\sum_{b=1}^B \text{I}_{i}^b},
\end{align*}
where $\text{I}_i^b$ is an indicator of whether the $i$-th observation is
\textit{not} included in the $b$-th bootstrap sample, and $\widehat{y}_i^{*b}$ is
the prediction obtained for observation $i$ from the $b$-th model fit. The
formula above is from the paper by \citet{efron1997improvements}, an alternative
formulation of the statistic is
\begin{align*}
  \widehat{\text{Err}^{(1)}} = \frac{\sum_{i=1}^n\sum_{b=1}^B \text{I}_{i}^b
    \text{L}(y_i, \hat{y}_i^{*b})}
  {\sum_{i=1}^n\sum_{b=1}^B \text{I}_{i}^b},
\end{align*}
found in the paper by \citet{efron1983estimating}. According to
\citet{efron1997improvements} these will be close for larger values of $B$, 
The bootstrap statistic we will use is based on the latter, and is given by
\begin{align*}
  \widehat{\mathcal{L}}^{fraud}_{BOOT} =
  \sum_{i=1}^n\frac{\sum_{b=1}^B \text{I}_{i}^b
    (1 - y_i)\hat{y}_i^{*b}}
  {\sum_{b=1}^B\sum_{i=1}^n \text{I}_{i}^b\hat{y}_i^{*b}}.
\end{align*}

%% file: simulating_data.tex
In order to study how well the method for selecting the model complexity 
based on the fraud loss works, we will perform a simulation study. First, we
examine and compare different setups of the selection approach. Subsequently,
we make a comparison to the most relevant alternative approach, based on the
AUC.

\subsection{Generating data}
\noindent
We want the synthetic datasets to possess many of the same characteristics as
real datasets from fraud detection applications, such as the dataset from the
Norwegian Tax Administration (Skatteetaten), that we study in Section
\ref{sec:realdat}. The common traits that we want to replicate, at least to some
degree, are correlated covariates, with margins of different types, some
continuous, some discrete, and an imbalance in the marginal distribution of the
outcome.

In order to draw the covariates, we follow a procedure that can be described in
a few stages. First, we draw a sample of a random vector, from a multivariate
distribution with uniform margins, i.e., a copula
\citep{nelsen2007introduction}. After simulating observations from the copula,
each of the margins is transformed to one of the distributions listed in Table
\ref{tab:margs}. The copula we will use is the $t_{2, \mathbf{R}}$-copula. Data are
then simulated by drawing from a multivariate $t_{2, \mathbf{R}}$-distribution,
i.e., a standard $t$-distribution with a $p\times p$ correlation matrix
$\mathbf{R}$ and $2$ degrees of freedom. Then, the margins are transformed via
the (univariate) $t_2$-quantile function, which makes the margins uniform, but
still dependent.

We specify the correlation matrix $\mathbf{R}$ of the multivariate
$t$-distribution by drawing a matrix from a uniform distribution over all
positive definite correlation matrices, using the algorithm described by
\citet{joe2006generating}. When looking at comparisons across a number of
different datasets, we always keep the correlation matrix $\mathbf{R}$ fixed.
Setting the correlation matrix by simulating it via an algorithm is just a
pragmatic way of specifying a large correlation matrix, while ensuring that it
is positive semidefinite. As for the correlation matrix, the distribution for
each margin is also drawn randomly from a list, but these distributions are also 
kept fixed, whenever we consider comparisons across different datasets.

\begin{table}
    \caption{List of the 17 different marginal distributions used to simulate
    covariates. \label{tab:margs}}
    \centering
    \input{marginal_distributions.tex}
\end{table}

Given the simulated covariates, probabilities $p(\mathbf{x})$ are computed,
and the binary outcomes $y_i, \, i=1,\dots,n,$ are simulated from
$\text{Bernoulli}\left(p(\mathbf{x}_i)\right) \,\text{distributions, where }
i=1, \dots,n.$ The probabilities follow the form
$$p(\mathbf{x}) = \frac{\exp(\beta_0 + f(\mathbf{x}))}
{1 + \exp(\beta_0 + f(\mathbf{x}))},$$
where the model is either linear, or an additive model (see Section 
\ref{sec:mods}). Unless otherwise stated, we simulate datasets with $p=100$
covariates. When simulating from a model with a linear predictor, we let $15$ of
the $100$ covariate effects be non-zero, and let these take values in the range
$(-0.77, 0.62),$ with an average absolute value of $0.34.$

In order to specify a
tree model, we draw a covariate matrix $\mathbf{\tilde{X}}$, and a response 
$\tilde{Y}$ to be used only for constructing the trees. The response is given by
$\tilde{Y}=BE_1 - (1 - B)E_2,$ where $B$ is a $\text{Bernoulli}(0.5)$ variable, and $E_1,$
$E_2$ are exponentially distributed with parameters $\alpha_1= 0.2$ and
$\alpha_2 = 0.1$, respectively. We then fit an additive tree model to this,
based on $15$ of the $100$ covariates. The resulting model is used to generate
datasets, keeping the model fixed across the different datasets. 

The parameter $\beta_0$ can change from dataset to dataset, as the average 
probability $p_0$ should be kept fixed. To achieve this, we solve the equation
$$\frac{1}{n}\sum_{i=1}^np(\mathbf{x}_i)=p_0$$ numerically, given the model
$f(\mathbf{x})$ and the covariate values of all the observations in the dataset.

%% file: marginal_distributions.tex
\begin{tabular}{c|c|c}
  \hline
  Family & Parameter & Value \\
  \hline
  Bernoulli & $p$ & 0.2 \\
  \hline
  Bernoulli & $p$ & 0.4 \\
  \hline
  Bernoulli & $p$ & 0.6 \\
  \hline
  Bernoulli & $p$ & 0.8 \\
  \hline
  Beta & $(\alpha, \beta)$ & $(1, 2)$\\
  \hline
  Beta & $(\alpha, \beta)$ & $(2, 1)$\\
  \hline
  Beta & $(\alpha, \beta)$ & $(2, 2)$\\
  \hline
  Gamma & $(\alpha, \beta)$ & $(1, 3)$\\
  \hline
  Gamma & $(\alpha, \beta)$ & $(3, 1)$\\
  \hline
  Gamma & $(\alpha, \beta)$ & $(3, 3)$\\
  \hline
  Normal & $(\mu, \sigma)$ & $(0, 1)$\\
  \hline
  Student's t & $\nu$ & $3$\\
  \hline
  Student's t & $\nu$ & $4$\\
  \hline
  Student's t & $\nu$ & $6$\\
  \hline
  Poisson & $\lambda$ & 1\\
  \hline
  Poisson & $\lambda$ & 3\\
  \hline
  Poisson & $\lambda$ & 5\\
  \hline
\end{tabular}

%% file: simulation.tex
\subsection{Data drawn from a logistic regression model}
\noindent
We first consider penalised logistic regression. We draw two datasets from a
logistic regression model, one for estimation and one for testing, by the method
described in Section \ref{subsec:simdata}, both with $n=1000$ observations. The 
coefficient $\beta_0$ is set so that the sample mean of $p(\mathbf{x})$ is $0.2$, 
which is quite high compared to a typical fraud detection setting, but is close to 
the average outcome in the dataset that we will discuss in Section \ref{sec:realdat}. 
In order to select the value of the regularisation parameter $\lambda$, we compare
bootstrap validation, and repeated cross validation with $10, 5, 3,$ or $2$
folds. We also try drawing the folds stratified on the outcome, so that each
fold has the same proportion of positive outcomes as the training sample. All
the experiments are repeated for $S=100$ simulated datasets.

In order to make the comparison fair, we balance the number of bootstrap folds
and the number of repetitions for the cross validation, so that the
computational complexity is roughly the same, assuming that the computational
complexity of fitting a model is proportional to the number of observations. All
validation procedures are adjusted so that they are comparable to 10-fold cross
validation without repetition. Since this involves fitting models to $10$
different datasets of $.9$ times the size of the total, we use $9$ folds in the
bootstrap validation, and $2,$ $4,$ and $9$ repetitions of $5$-, $3$-, and
$2$-fold cross validation, respectively. We also do the same, with double the
number of repetitions across 10-, 5-, 3- and 2- fold cross validation, and with
$18$ bootstrap folds. 

\begin{table}
  \caption{Relative fraud loss averaged over all values of $k$, for data simulated 
  	from the linear model with $p=100$ $n=1000$, and fitted with a linear model. \label{tab:tab1}}
  \input{sim_linear_linear.tex}
\end{table}

We define relative fraud loss, compared to the minimum in each simulation 
$s=1, 2, \dots, S,$ over the alternatives for the tuning parameter as
\begin{align}
  RFL(k) &= 
  \frac{\sum_{s=1}^S \sum_{i=1}^n
  \widehat {y_{i,s}}^{{\lambda_{sel}}}(k)(1 - y_{i,s}) /
  \sum_{i=1}^n\widehat {y_{i,s}}^{{\lambda_{sel}}}(k)}{
  \sum_{s=1}^S \sum_{i=1}^n  \widehat{y_{i,s}}^{{\lambda_{opt}}}(k)(1 - y_{i,s}) /
  \sum_{i=1}^n\widehat {y_{i,s}}^{\lambda_{opt}}(k)},
\label{rel_fraud_loss}
\end{align}
for a given value of $k$. Here, $\widehat {y_{i,s}}^{{\lambda_{sel}}}(k)$ is the
prediction of $Y_{i,s}$ from the model resulting from a particular choice of tuning
parameter, for a given $k$, and $\widehat {y_{i,s}}^{\lambda_{opt}}(k)$ is the
corresponding prediction from the model that is optimal, over all values of the tuning
parameter, for that simulation $s$. This is computed for 
$k = 10, 20, \dots, 980, 990,$ and the average 
$\frac{1}{99}\sum_{j=1}^{99}RFL(k_{j})$ over all values $k_{j}$ of $k$ is reported
in Table \ref{tab:tab1}. As expected, doubling the number of repetitions leads
to better performance. Looking at the different validation schemes, it seems
like there is a tendency that cross validation with 2 or 3 folds is the better
option, and that there is a slight advantage to stratification, but this does
not seem very conclusive. Based only on these results, we would therefore
suggest using 2-fold cross validation, and to stratify on $y$ at least if there
are so few observations where $y=1,$ that there is a risk of getting folds where
all observations have a negative outcome.

\begin{figure}
  \includegraphics[width=400pt]{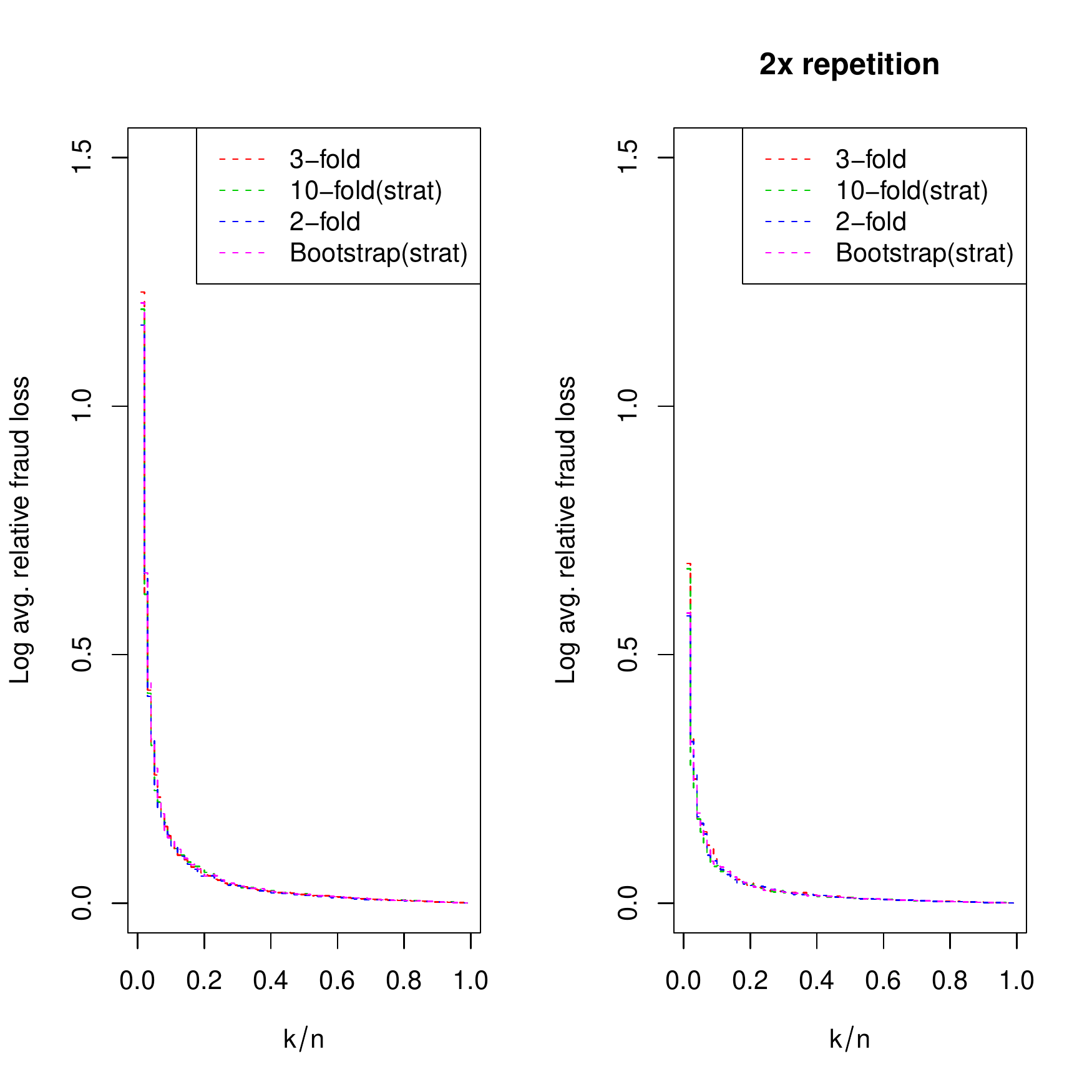}
  \caption{Plot of the log relative fraud loss as a function of the fraction
  $\tau = k/n$ of the observations that are selected, for a selection of the
  methods for setting the tuning parameter. The logistic regression model was used
  both to simulate the data, and to make predictions. \label{fig:fig1}}
\end{figure}

In Figure \ref{fig:fig1}, the logarithm of the relative fraud loss is plotted as a function 
of the proportion $k/n$ of cases that are selected, for a selection of the validation 
procedures. From these, we see that there is not a very large difference between the 
different types. Further, it is hardest to select the best cases for lower values of 
$k/n$, as expected.

\begin{figure}
  \includegraphics[width=400pt]{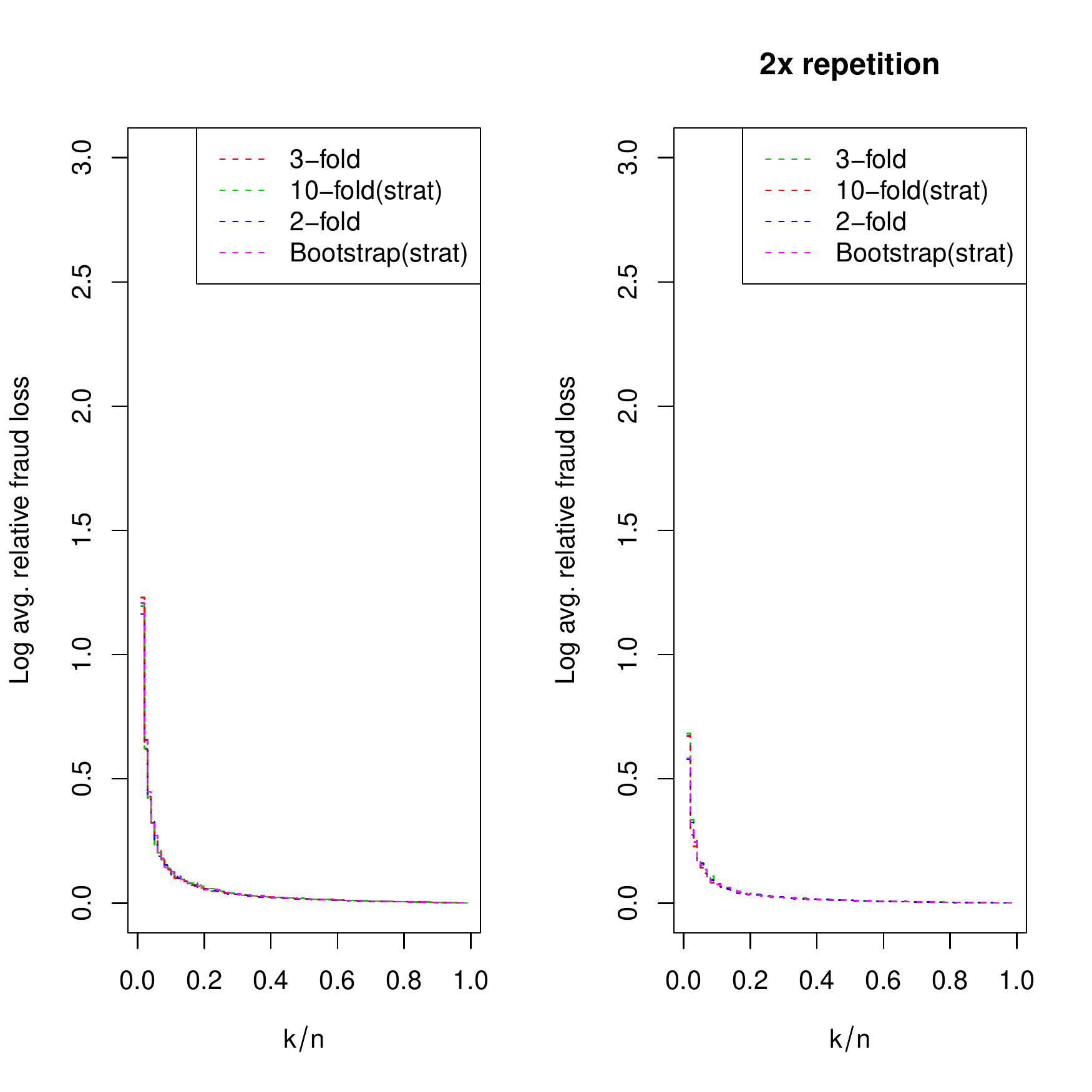}
  \caption{Plot of the relative fraud loss as a function of the fraction $\tau = k/n$
    of the observations that are selected, for a selection of the methods for 
    setting the tuning parameter. The logistic regression model was used
    both to simulate the data, and the additive tree model to make predictions.
    \label{fig:fig2}}
\end{figure}

We repeat this experiment for the same datasets, but instead of estimating the
probabilities using a penalised logistic regression model, we use an additive tree 
model fitted by boosting. The penalty parameter to be selected, is now the total 
number $M$ of components of the additive tree model. The average relative fraud 
loss for the different ways of choosing the number of iterations is reported in Table 
\ref{tab:tab2}, and the logarithm of the relative fraud loss is plotted as a function of 
$k/n$ in Figure \ref{fig:fig2}. Compared to the ridge regression fits, the relative fraud 
loss is now greater, meaning that the selected model differs more in size, compared to 
the minimum for each value of $k$. The best alternative now seems to be the 
bootstrap variants, with stratified 3-fold cross validation being the closest contender.

\begin{table}
  \caption{Relative fraud loss averaged over all values of $k$, for data simulated 
  	from the linear model with $p=100$ $n=1000$, and fitted with an additive tree model. \label{tab:tab2}}
  \input{sim_linear_trees.tex}
\end{table}

\subsection{Data drawn from an additive tree model}
  \noindent
  Next, we do a similar experiment, the difference being the model that the data are drawn from. Instead of a logistic regression model, we draw from a model where the 
  linear predictor is replaced with an additive tree model, as previously outlined. We 
  first estimate models using penalised logistic regression.
  \begin{table}
    \caption{Relative fraud loss averaged over all values of $k$, for data simulated 
    	from the additive tree model with $p=100$ $n=1000$, and fitted with the linear model.
    	\label{tab:tab3}}
    \input{sim_trees_linear.tex}
  \end{table}
  The results of this are summarised in Table \ref{tab:tab3}.
  \begin{table}
    \caption{Relative fraud loss averaged over all values of $k$, for data simulated 
    	from the additive tree model with $p=100$ $n=1000$, and fitted with the additive tree model.\label{tab:tab4}}
    \input{sim_trees_trees.tex}
  \end{table}
  We also estimate additive tree models, the results for which are summarised in
  Table \ref{tab:tab4}. In the first case, 2-fold cross-validation gave the best results, with stratified 3-fold cross validation in second place. Curiously, the error does not 
  seem to be smaller when doubling the number of repetitions, which could suggest 
  that it is easier to select the best model for the data simulated from a more complex 
  model. For the latter case, 2-fold cross validation seems to be the best option, 
  based on the average relative fraud loss. It could be argued that real data are most 
  likely to follow a model that is more complex than a logistic regression model with a 
  \textit{linear} predictor, and therefore that repeated 2-fold cross validation is the 
  more reliable option, overall.

\subsection{Comparison with an alternative approach}
  \noindent
  Next, we want to compare our approach to other relevant methods. One such 
  method is to set the penalty parameter using the AUC as a criterion. This is a 
  popular measure for assessing the performance of a binary regression model in
  terms of discrimination. It is also related to ranking, and is therefore a natural 
  alternative to fraud loss. In fact, the AUC can on a population level be seen to be 
  equivalent to the probability that an observation where $Y = 0$ will be given a lower
  probability than one where $Y=1$. Hence, if one model has a higher AUC than
  another, then the aforementioned probability will be highest for the model
  with the highest AUC \citep{clemenccon2008ranking}. Symbolically, this can be
  written as
  \begin{align*}
    \text{AUC}\left(\hat{p}\right) &= P\left(\hat p(x_i) \geq \hat p(x_j)
    \vert Y_i = 1, Y_j = 0\right),
  \end{align*}
  which may be estimated by the Wilcoxon type statistic
  \begin{align*}
  \widehat{\text{AUC}} = \frac{\sum_{i=1}^n\sum_{j=1}^n(1 - y_i)y_j
  	\mathcal{I}\left(\hat{p}(\mathbf{x}_j) > \hat{p}(\mathbf{x}_i)\right)}
  {\sum_{i=1}^ny_i\sum_{i=1}^n(1 - y_i)}.
  \end{align*}
  \begin{table}
    \caption{Relative fraud loss averaged over $k$ for the methods based on the 
    	AUC and the fraud loss, for data simulated with $p=100$ $n=1000$.
      \label{table:aucVsfraud_ngreaterthanp}}
    \input{vs_auc_n_greaterthan_p.tex}
  \end{table}
 
  The log-likelihood function, or likelihood-based measures, such as the Akaike 
  information criterion (AIC), are also commonly used to set tuning parameters, but 
  we will here disregard these. However, they are not particularly relevant for the 
  problem, as we are not interested in finding the model that gives the best fit to all 
  the data. Further, the log-likelihood function often explodes numerically when 
  some of the probabilities become very close to $1$. In our experiments, this 
  happened often during cross validation when we evaluated the log-likelihood 
  function for the data that were not used for estimation.

  The first simulations are based on the same models as previously discussed,
  using 2-fold cross validation with and without repetition in both methods. Table
  \ref{table:aucVsfraud_ngreaterthanp} reports the resulting average fraud loss.
  The average over a smaller range of values of $k/n=0.16, 0.17, \dots, 0.25,$ 
  which is realistic in practice for the fraud setting is also shown in the table. When
  looking at the average over $k/n$ from $0.01$ to $0.99$, it seems that it is
  advantageous to select the model complexity via the cross validated fraud loss
  when estimating an additive tree model, whatever the data generating model is.
  When we only consider an aggregate over values of $k/n$ from $0.16$ to $0.25$, 
  we see the same, and in addition there also seems to be a slight benefit to
  using  the fraud loss, when fitting penalised logistic regression models to data simulated from an additive tree model.
  
\begin{figure}
    \includegraphics[width=400pt]{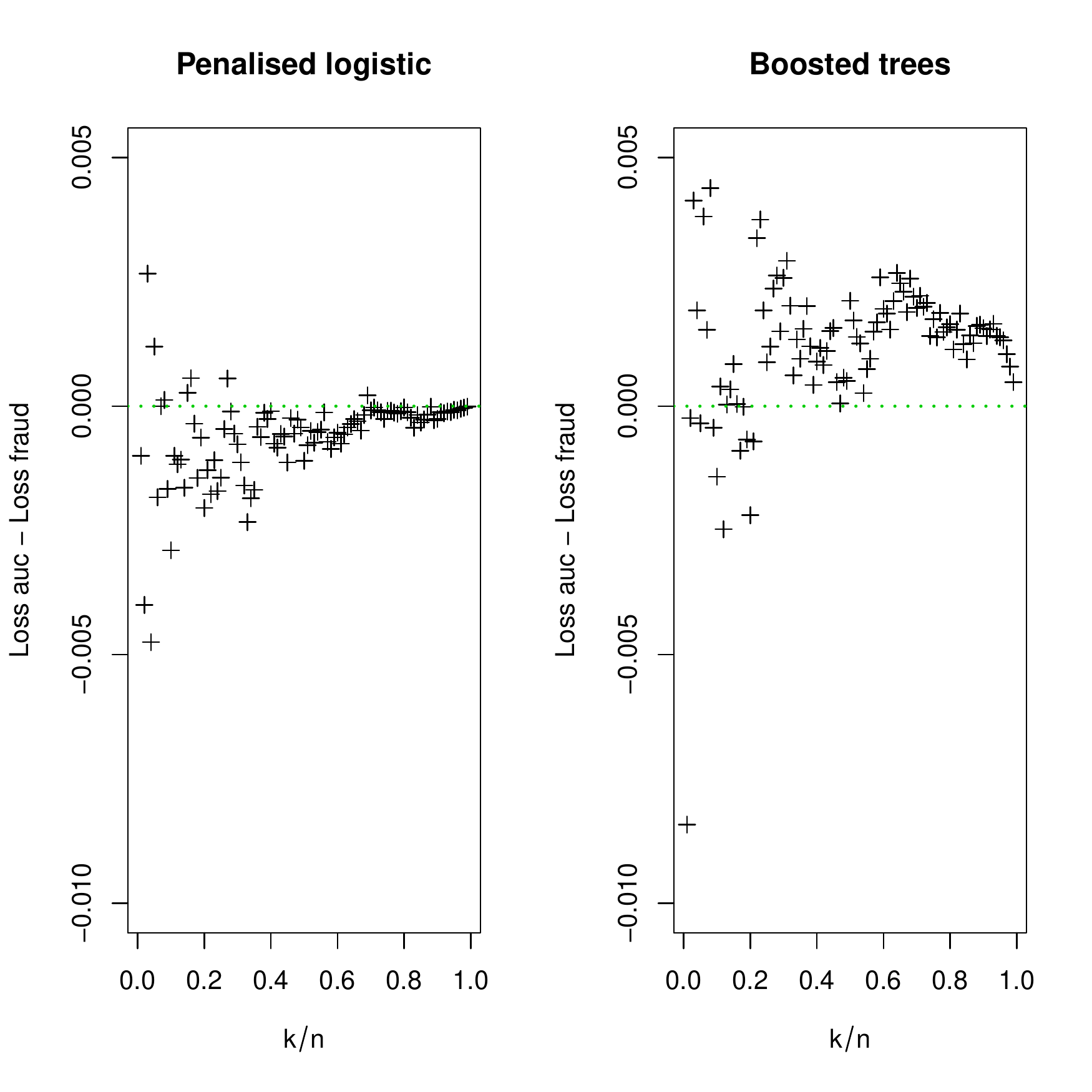}
    \caption{Plot of the difference in fraud loss when selecting the model complexity
      according to the AUC and the fraud loss, respectively. The data are simulated 
      from the logistic regression model with $n=1000$ and $p=100$. 
    \label{fig:vs_auc_n_gt_p}}
  \end{figure}

\begin{figure}
    \includegraphics[width=400pt]{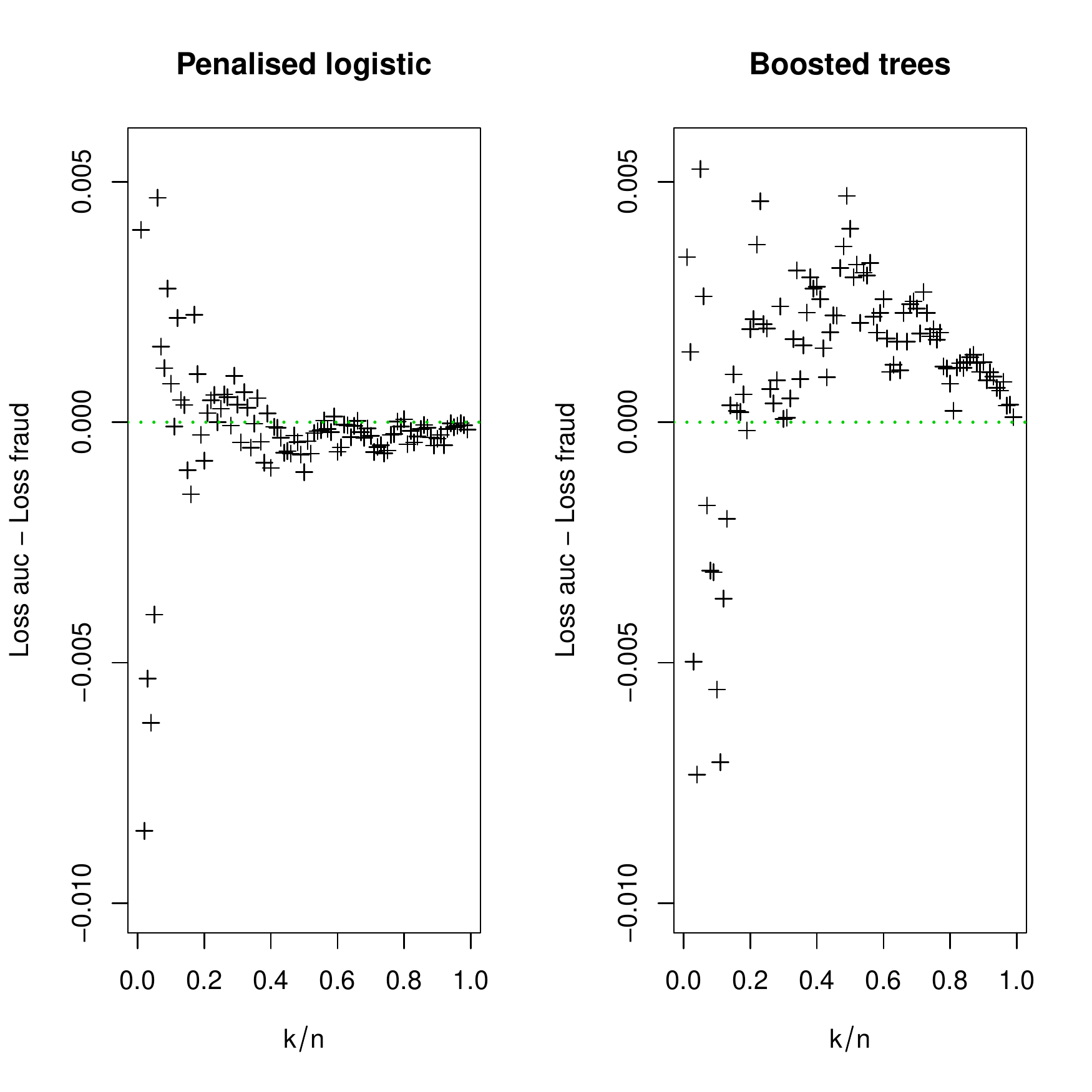}
    \caption{Plot of the difference in fraud loss when selecting the model complexity
    	according to the AUC and the fraud loss, respectively. The data are simulated 
    	from additive tree model with $n=1000$ and $p=100$.
    \label{fig:vs_auc_n_gt_p_2}}
  \end{figure}

  In Figure \ref{fig:vs_auc_n_gt_p}, the average difference in fraud loss for the two 
  approaches applied to the data simulated from logistic regression models, is 
  plotted as a function of $k/n.$ It seems that in neither case, one or the other 
  method is consistently better across the entire grid over $k/n$. However, there is 
  a tendency, especially when $k/n$ is larger than roughly $0.2$, that the fraud loss 
  works best when estimating an additive tree model, but not when estimating a 
  penalised logistic regression model. Figure \ref{fig:vs_auc_n_gt_p_2} is a similar plot, 
  but where the data are simulated from the additive tree model. It seems, in this case, 
  that the fraud loss is more favourable when a penalised logistic regression model is 
  estimated, compared to when the data were simulated from the logistic regression 
  model.

  
  We repeat the experiment, but we now expand the datasets so that the number of
  covariates is $p=n=1000.$ These are simulated in the same way as for
  $p=100.$ We again simulate data both from a logistic regression model, and
  from an additive tree model, and scale the number of covariates that the
  response depends on with the dimension of the covariate matrix, so that it in both
  cases depends on $150$ covariates. For the logistic regression model, the $150$ 
  non-zero effects take values in the interval $(-0.67, 0.85),$ and the average 
  absolute value of these is $0.198.$ A comparison of the two  approaches for 
  selecting the model complexity, in all four combinations of data-generating and 
  estimated model, is summarised in terms of the average relative fraud loss in Table 
  \ref{table:aucVsfraud_neqp}. The average relative fraud loss over the whole 
  range of $k/n$ is now lowest when using the approach based on the fraud loss, 
  except when both the data-generating and the estimated model are logistic. When 
  we only look at the average over $k/n=0.16,\dots, 0.25,$ it seems to be beneficial 
  to select the model complexity with the fraud loss in all cases, although the
  difference between the results from the two methods is quite small when estimating 
  a penalised regression model.

\begin{figure}
    \includegraphics[width=400pt]{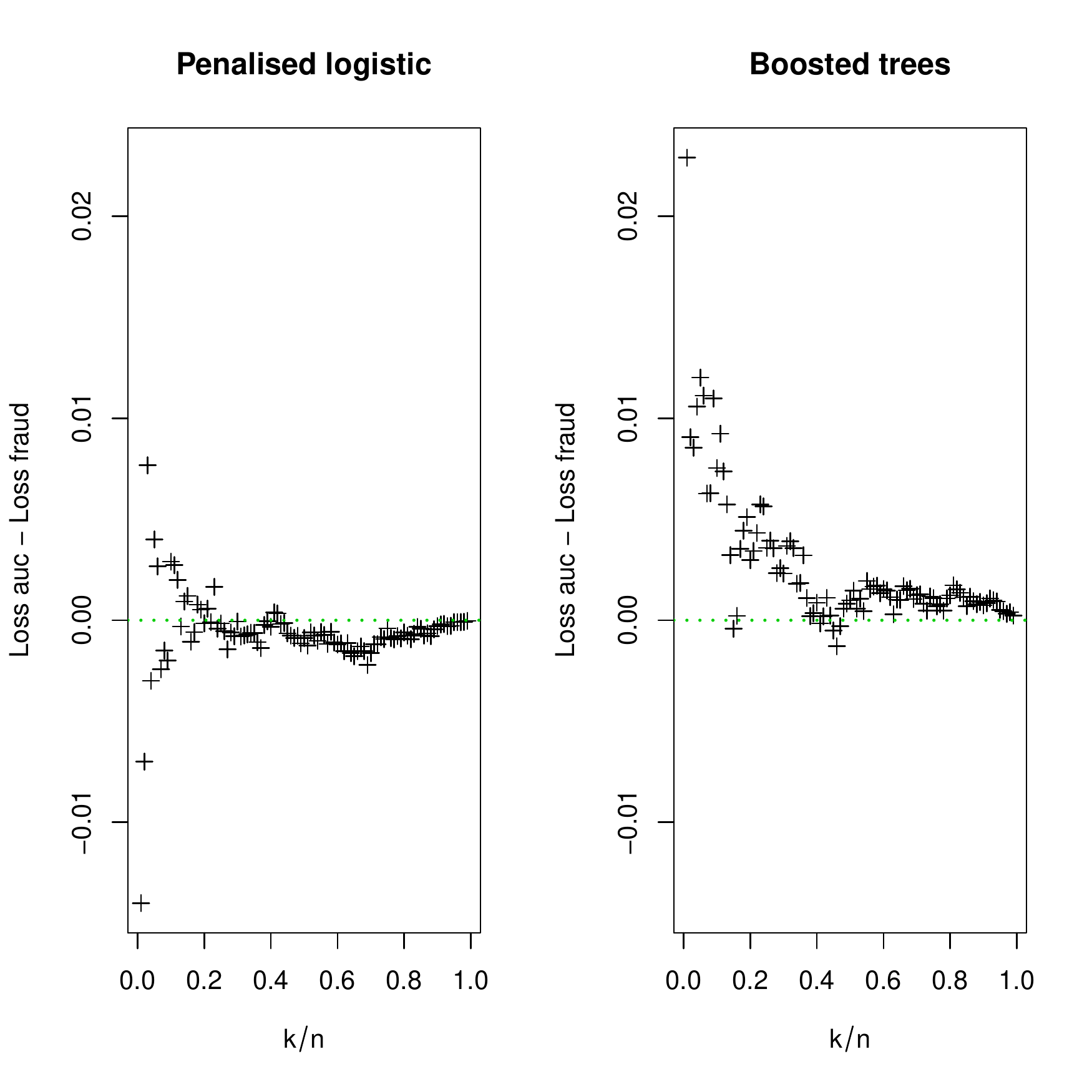}
    \caption{Plot of the difference in fraud loss when selecting the model complexity
    	according to the AUC and the fraud loss, respectively. The data are simulated 
    	from the logistic regression model with $n=1000$ and $p=1000$. 
      covariates. \label{fig:vs_auc_n_eq_p}}
  \end{figure}

\begin{figure}
    \includegraphics[width=400pt]{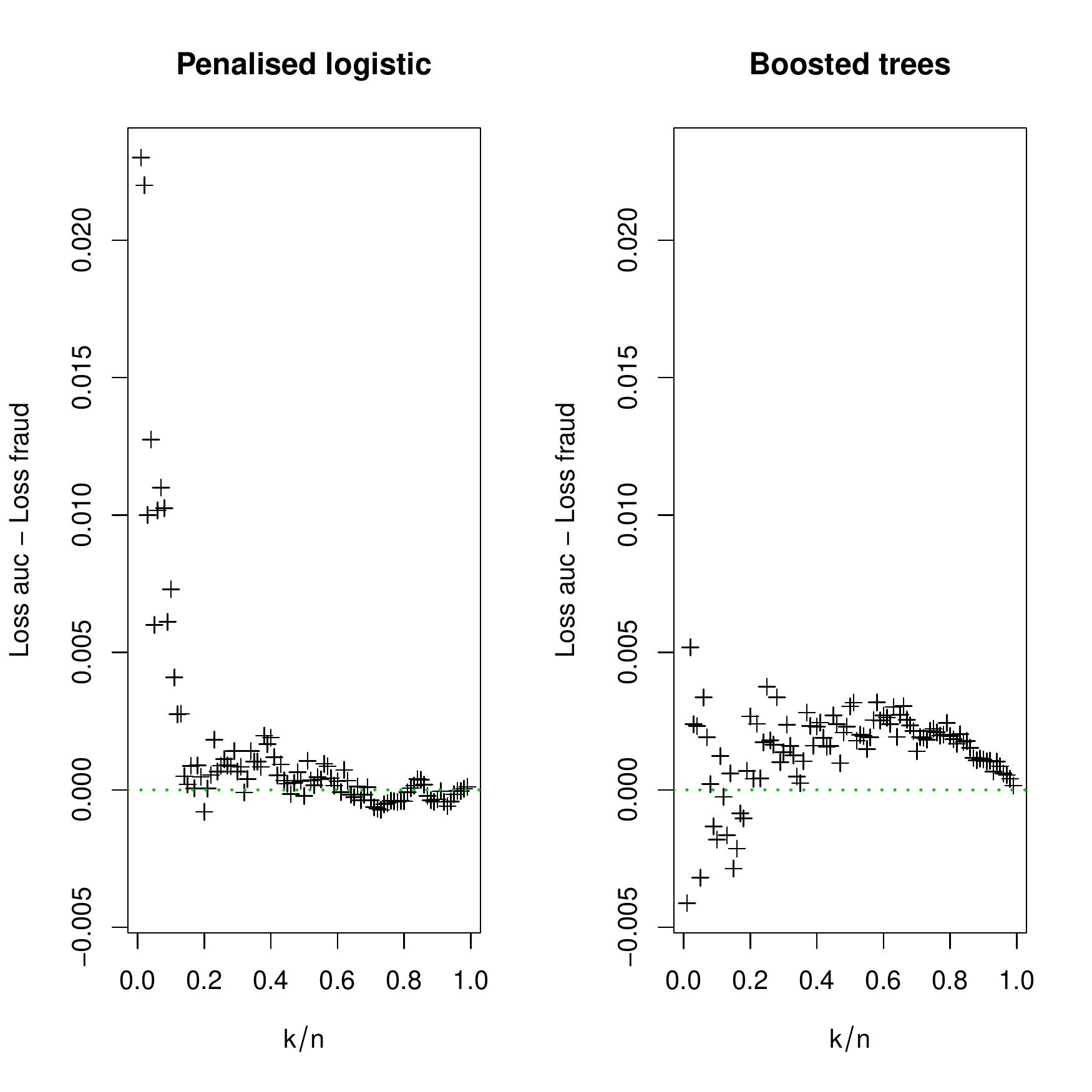}
    \caption{Plot of the difference in fraud loss when selecting the model complexity
    	according to the AUC and the fraud loss, respectively. The data are simulated 
    	from the additive tree model with $n=1000$ and $p=1000$.
      covariates. \label{fig:vs_auc_n_eq_p_2}}
  \end{figure}

  Figure \ref{fig:vs_auc_n_eq_p} is a plot corresponding to Figure
  \ref{fig:vs_auc_n_gt_p}, but for $p=1000.$ The fraud loss is now lowest
  overall, when the tuning parameter is selected with the cross validated fraud
  loss for the boosted models, but not for the penalised logistic regression models. 
  When the data are simulated from an additive tree model, as shown in Figure 
  \ref{fig:vs_auc_n_eq_p_2}, there seems to be an advantage to using the fraud loss
  when estimating penalised regression models, at least for $k/n$ up to $0.2$. Fraud 
  loss also seems to be best for the boosted tree models, perhaps except when 
  $k/n < 0.2$, possibly due to a high variance. These results could indicate that it is 
  better to chose the penalty parameter by cross validating the fraud loss, than by 
  the AUC, when the model is misspecified. 

  \begin{table}
    \caption{Relative fraud loss averaged over $k$ for the methods based on the 
    	AUC and the fraud loss, for data simulated with $p=1000$ $n=1000$.
      \label{table:aucVsfraud_neqp}}
    \input{vs_auc_n_eq_p.tex}
  \end{table}

  \begin{table}
    \caption{Relative fraud loss averaged over $k$ for the methods based on the 
    	AUC and the fraud loss, for data simulated with $p=4000$ $n=1000$.
      \label{table:aucVsfraud_pgrn}}
    \input{vs_auc_p_greaterthan_n.tex}

  \end{table}

  \begin{figure}
    \includegraphics[width=400pt]{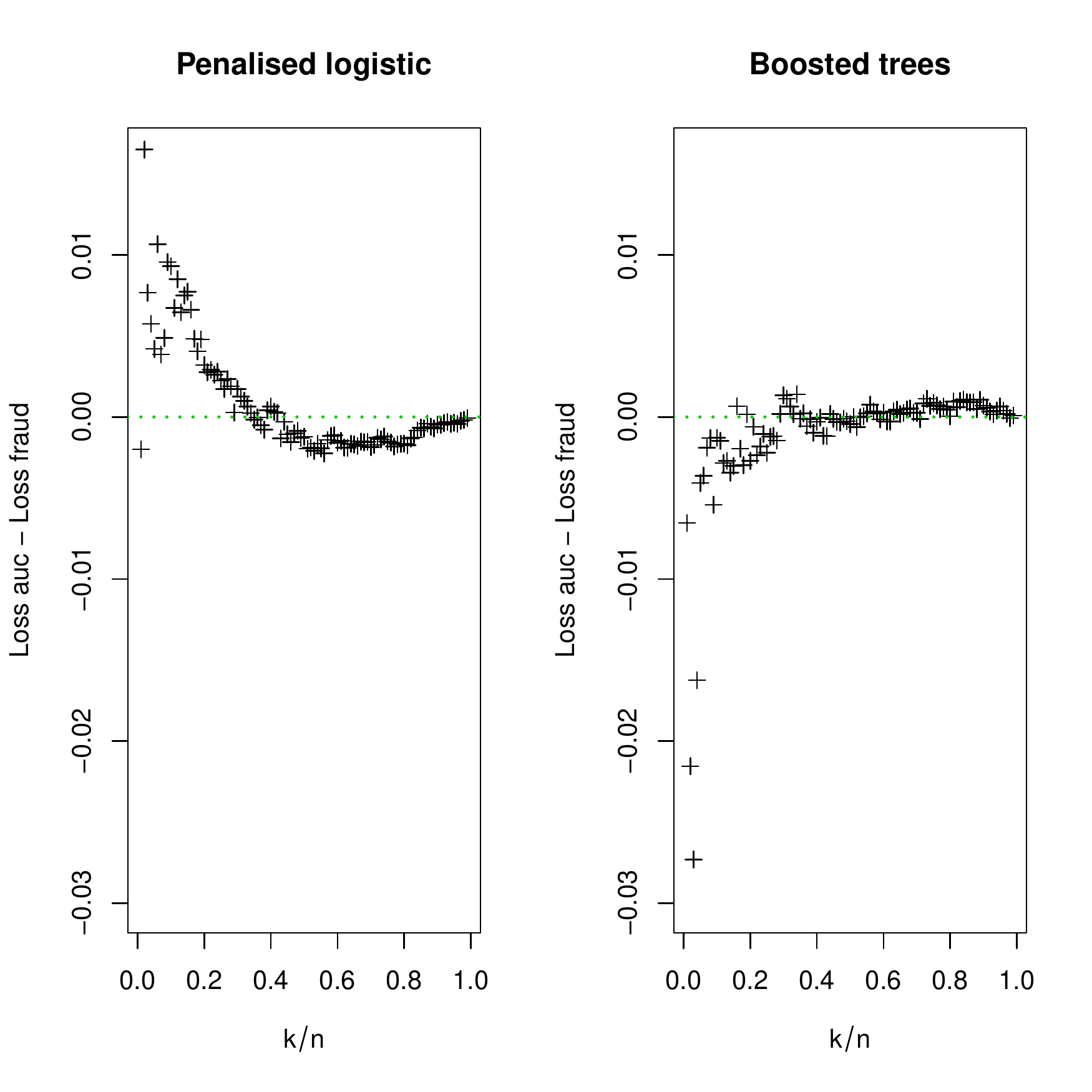}
    \caption{Plot of the difference in fraud loss when selecting the model complexity
    	according to the AUC and the fraud loss, respectively. The data are simulated 
    	from the logistic regression model with $n=1000$ and $p=4000$.
      covariates. \label{fig:vs_auc_p_gr_n}}
  \end{figure}

  \begin{figure}
    \includegraphics[width=400pt]{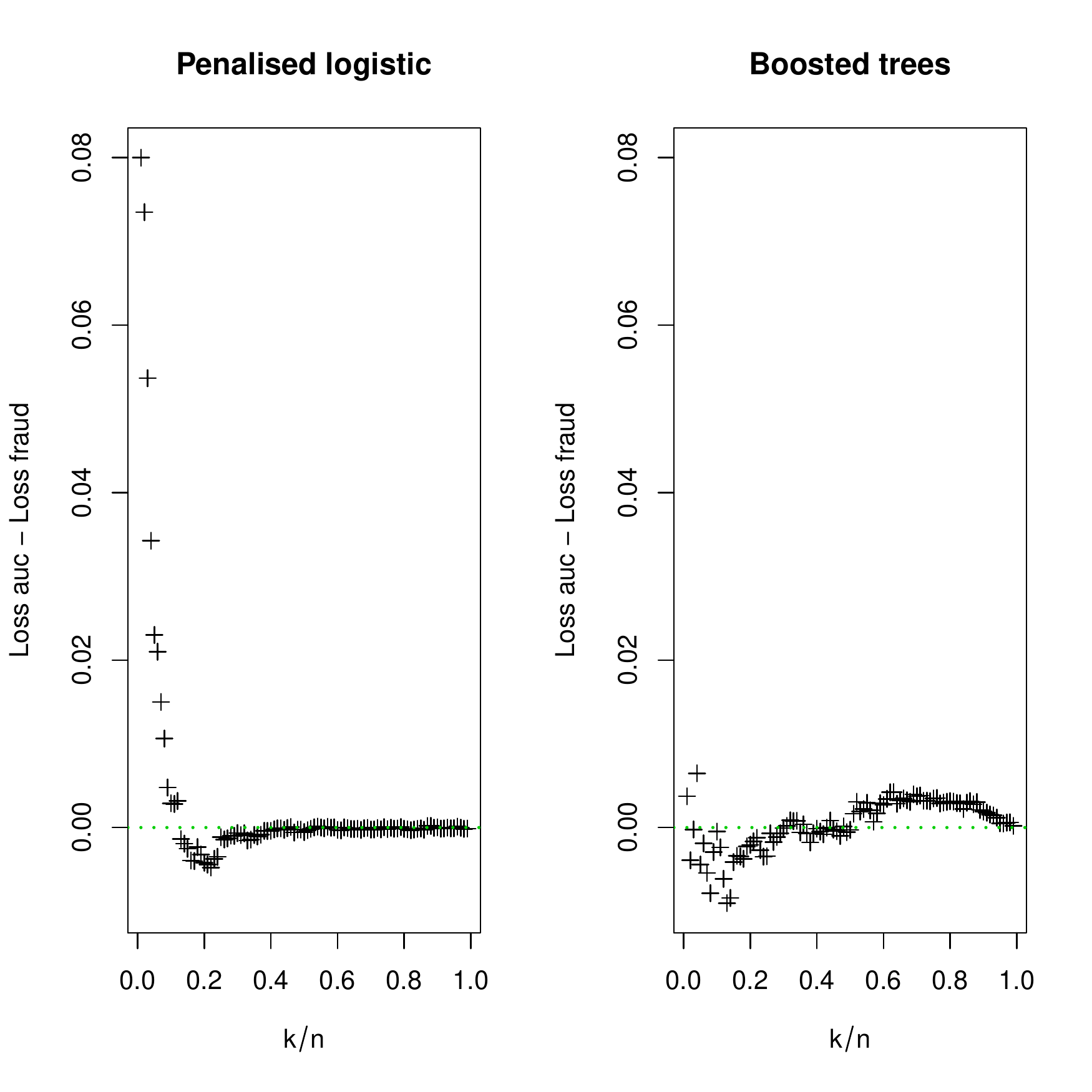}
    \caption{Plot of the difference in fraud loss when selecting the model complexity
    	according to the AUC and the fraud loss, respectively. The data are simulated 
    	from the additive tree model with $n=1000$ and $p=4000$.
      covariates. \label{fig:vs_auc_p_gr_n}}
  \end{figure}

  We repeat the experiment again, now in a context where $p > n$. Specifically,
  we let $p=4000,$ while we keep the number of observations at $n=1000.$
  With the exception of the correlation matrix $R$, the model parameters are scaled 
  up in the same way as when the number of covariates was changed from $100$ to 
  $1000$. The correlation matrix is now constructed from one $(1000) \times (1000)$ 
  correlation matrix that is stacked diagonally, giving a block-diagonal correlation 
  matrix where $4\times {10}^{6}$ out of the $16\times{10}^{6}$ entries are non-zero. 
  When the data are simulated from the logistic regression model, we see in Figure 
  \ref{fig:vs_auc_p_gr_n} that picking the tuning parameter for a penalised logistic 
  regression model using the fraud loss, generally works better than the AUC, at least 
  when $k/n < 0.4,$ and that for the boosted trees it seems to be consistently worse, 
  or at least not better. For the data drawn from an additive tree model, the fraud loss 
  is better for values of $k/n$ up to about $0.12$ for the penalised logistic regression 
  models, with more or less no difference for $k/n > 0.4.$ For the boosted trees, the 
  fraud loss performs somewhat worse for values of $k/n$ up to about $0.3.$ All of 
  the simulations are summed up in Table \ref{table:aucVsfraud_pgrn}, demonstrating 
  that the fraud loss on average performs better than the AUC for the penalised 
  logistic regression models, regardless of whether the data are simulated from a 
  logistic regression model or an additive tree model when we look at the entire range 
  of $k$. If we only look at $k/n=0.16, 0.17, \dots, 0.25$, the fraud loss performs 
  better only for penalised logistic regression when the data are drawn from a logistic
  regression model. 

  All in all, selecting the model complexity by cross validating the fraud loss seems 
  to work quite well, compared to using the AUC, especially for values of $k/n$ close 
  to the marginal probability of $Y$. This is good, as one could argue that these are 
  the most interesting values in most fraud detection applications. The fraud loss 
  does not outperform the AUC in all cases, such as for the boosted tree models for 
  $p > n$, or the penalised regression models when the data are simulated from a 
  logistic regression model, and $p < n$. A possible explanation for the first case 
  could be that when estimating the probabilities is difficult due to $p>n,$ and the 
  data generating model is complex, then trying to adapt the model locally to $k/n$,
  as in the method using the fraud loss, could introduce instability. For the latter case, 
  it could be that the estimation problem is so simple that one more or less recovers 
  the data generating model, and that there might not be too much to gain from 
  adapting the model to a specific $k/n$.

%% file: sim_linear_linear.tex
\begin{tabular}{r| c | c | c | c | c }
  \hline
  Notes & 10-fold cv & 5-fold cv & 3-fold cv & 2-fold cv & Bootstrap \\
  \hline
  &1.0764 & 1.0766 & 1.0753 & 1.0725 & 1.0763\\
  \hline
  stratified & 1.0745 & 1.0767 & 1.0730 & 1.0743 & 1.0768\\
  \hline
  2x repeat & 1.0385 & 1.0366 & 1.0364 & 1.0366 & 1.0389\\
  \hline
  2x repeat, stratified & 1.0396 & 1.0384 & 1.0355 & 1.0354 & 1.0366\\
  \hline
\end{tabular}

%% file: sim_linear_trees.tex
\begin{tabular}{r| c | c | c | c | c }
  \hline
  Notes & 10-fold cv & 5-fold cv & 3-fold cv & 2-fold cv & Bootstrap \\
  \hline
  & 1.2060 & 1.1779 & 1.1617 & 1.1424 & 1.1578\\
  \hline
  stratified & 1.2123 & 1.1716 & 1.1534 & 1.1403 & 1.1496\\
  \hline
  2x repeat & 1.0586 & 1.0507 & 1.0526 & 1.0541 & 1.0464\\
  \hline
  2x repeat, stratified & 1.0552 & 1.0523 & 1.0499 & 1.0515 & 1.0472\\
  \hline
\end{tabular}  

%% file: sim_trees_linear.tex
\begin{tabular}{r| c | c | c | c | c }
  \hline
  Notes & 10-fold cv & 5-fold cv & 3-fold cv & 2-fold cv & Bootstrap \\
  \hline
  & 1.0190 & 1.0195 & 1.0193 & 1.0189 & 1.0194\\
  \hline  
  stratified & 1.0199 & 1.0195 & 1.0191 & 1.0190 & 1.0193\\
  \hline
  2x repeat & 1.0197 & 1.0193 & 1.0193 & 1.0190 & 1.0199\\
  \hline
  2x repeat, stratified & 1.0196& 1.0195 & 1.0192 & 1.0193 & 1.0196\\
  \hline
\end{tabular}

%% file: sim_trees_trees.tex
\begin{tabular}{r| c | c | c | c | c }
  \hline
  Notes & 10-fold cv & 5-fold cv & 3-fold cv & 2-fold cv & Bootstrap \\
  \hline
  & 1.0463 & 1.0450 & 1.0452 & 1.0454 & 1.0450\\
  \hline  
  stratified & 1.0469 & 1.0474 & 1.0457 & 1.0451 & 1.0454\\
  \hline
  2x repeat & 1.0467 & 1.0464 & 1.0456 & 1.0435 & 1.0451\\
  \hline
  2x repeat, stratified & 1.0473 & 1.0453 & 1.0464 & 1.0450 & 1.0452\\
  \hline
\end{tabular}

%% file: vs_auc_n_greaterthan_p.tex
\begin{tabular}{c|c|c|c|c}
  \hline
  Simulation model & Logistic & Logistic & Additive trees & Additive trees \\
  Estimation model & Logistic & Additive trees &  Logistic & Additive trees \\
  \hline
    \multicolumn{2}{c}{Average over all $K = 10, 20, \dots 990$:}\\
  \hline
  auc  &1.0335 &1.0763 &1.0186 &1.0482\\
  fraud  &1.0371 &1.0741 &1.0195 &1.0442\\
  auc, 2x repeat  &1.0337 &1.0730 &1.0183 &1.0470\\
  fraud, 2x repeat  &1.0350 &1.0721 &1.0186 &1.0454\\
  \hline
    \multicolumn{2}{c}{Average over $K = 160, 170, \dots 240, 250$:}\\
  \hline
  auc  &1.0349 &1.0626 &1.0228 & 1.0583\\
  fraud  &1.0356 &1.0591 &1.0232 & 1.0531\\
  auc, 2x repeat  &1.0338 &1.0610 &1.0228 & 1.0570\\
  fraud, 2x repeat  &1.0359 &1.0601 &1.0225 & 1.0545\\
  \hline
\end{tabular}

%



%% file: vs_auc_n_eq_p.tex
\begin{tabular}{c|c|c|c|c}
  \hline
  Simulation model & Logistic & Logistic & Additive trees & Additive trees \\
  Estimation model & Logistic & Additive trees &  Logistic & Additive trees \\
  \hline
  \multicolumn{2}{c}{Average over all $k = 10, 20, \dots 990$:}\\
  \hline
  auc              &1.0338 &1.0925 &1.0232 &1.0572\\
  fraud            &1.0352 &1.0919 &1.0218 &1.0537\\
  auc, 2x repeat   &1.0337 &1.0991 &1.0239 &1.0559\\
  fraud, 2x repeat &1.0351 &1.0910 &1.0216 &1.0539\\
  \hline
  \multicolumn{2}{c}{Average over $k = 160, 170, \dots 240, 250$:}\\
  \hline
  auc              &1.0394 &1.0875 &1.0287 &1.0729\\
  fraud            &1.0375 &1.0824 &1.0316 &1.0683\\
  auc, 2x repeat   &1.0389 &1.0920 &1.0301 &1.0705\\
  fraud, 2x repeat &1.0387 &1.0847 & 1.0294 &1.0693\\
  \hline
\end{tabular}

%% file: vs_auc_p_greaterthan_n.tex
\begin{tabular}{c|c|c|c|c}
  \hline
  Simulation model & Logistic & Logistic & Additive trees & Additive trees \\
  Estimation model & Logistic & Additive trees &  Logistic & Additive trees \\
  \hline
  \multicolumn{2}{c}{Average over all $k = 10, 20, \dots 990$:}\\
  \hline
  auc              &1.0415 &1.0526 &1.0229 &1.0453\\
  fraud            &1.0400 &1.0577 &1.0182 &1.0457\\
  auc, 2x repeat   &1.0412 &1.0520 &1.0229 &1.0443\\
  fraud, 2x repeat &1.0388 &1.0547 &1.0187 &1.0438\\
  \hline
  \multicolumn{2}{c}{Average over $k = 160, 170, \dots 240, 250$:}\\
  \hline
  auc              &1.0538 &1.0605 &1.0187 &1.0507\\
  fraud            &1.0498 &1.0655 &1.0223 &1.0585\\
  auc, 2x repeat   &1.0532 &1.0567 &1.0189 &1.0491\\
  fraud, 2x repeat &1.0468 &1.0591 &1.0239 &1.0530\\
  \hline
\end{tabular}

%% file: illustration.tex
In this section, we will consider a dataset of controlled cases of
potential VAT fraud from the Norwegian Tax Administration (Skatteetaten). The
data are sensitive. Therefore, they have been somewhat manipulated in
order to be anonymised, and very little meta-information, such as what the
covariates represent, is included. The data were collected in 39 different
2-month periods, and include in total around $n=50000$ observations of $p=555$
covariates, where $20$ of the covariates are binary, $12$ are categorical, and
$523$ are numerical. We recode the categorical covariates as binary variables,
which then effectively gives us $p=616$ covariates. There are some missing
values in the dataset. In order to be able to fit penalised logistic regression
models, we impute the median for the missing values of the numerical variables.
For the categorical variables, we instead recode them with an extra level that
corresponds to a missing observation.

\begin{figure}
  \includegraphics[width=400pt]{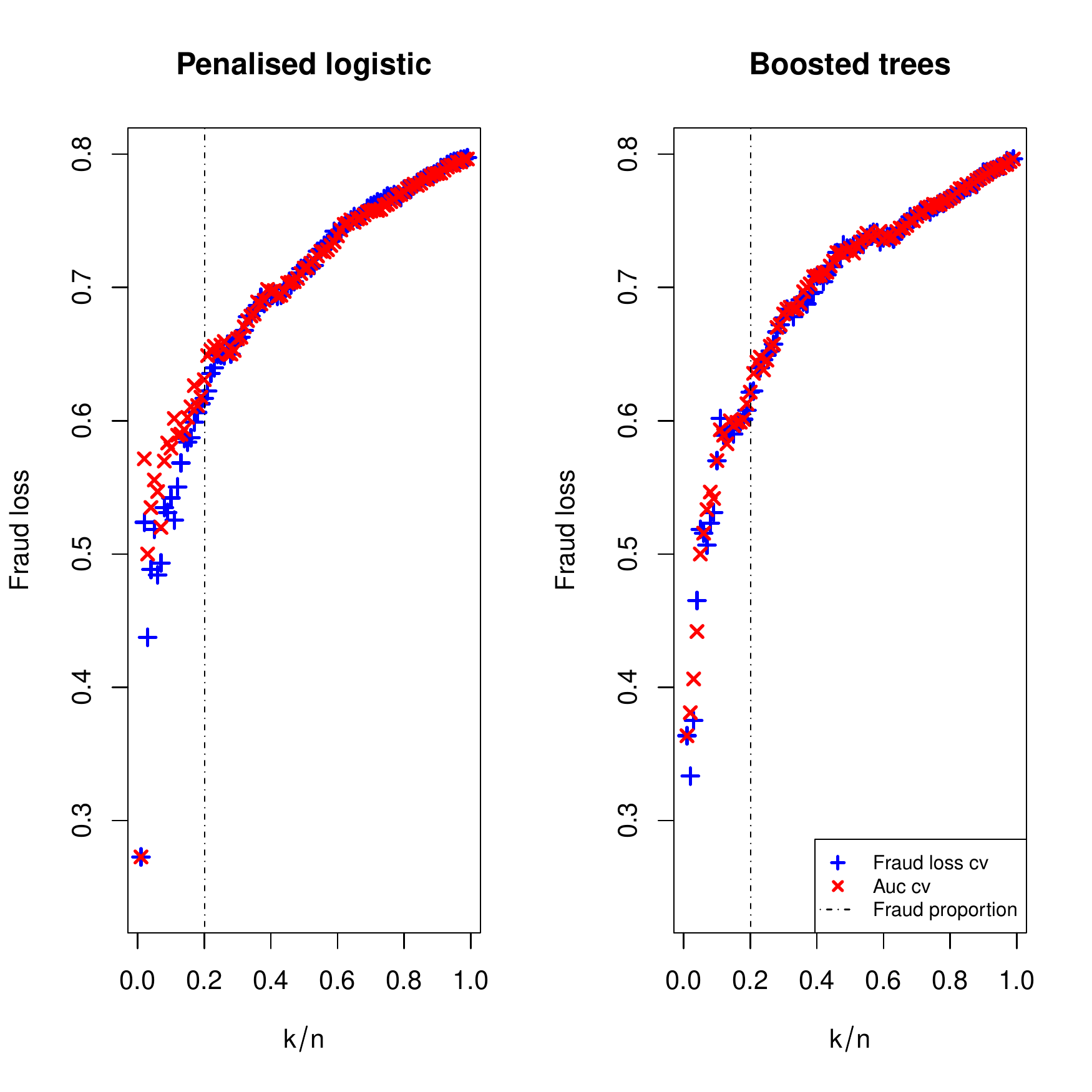}
  \caption{Plot of the fraud loss when selecting the model
    according to the AUC and fraud loss, respectively. Dataset of VAT fraud from
    the Norwegian Tax Administration. \label{fig:anls1}}
\end{figure}

This dataset serves as an example of one particular case, where selecting the top
cases is relevant. As an illustration, we take the data from the six 2-month
periods leading up to, but not including, the 12th 2-month period as a training
set, a total of $7843$ observations, out of which $1648$ are recorded as fraudulent. 
We use repeated 2-fold cross validation, with $9$ repetitions, to set the penalty parameter of a penalised logistic regression model, and the number of components 
in a boosted tree model. We set these parameters by cross validating the AUC and 
the fraud loss for $k/n = 0.01, 0.02, \dots, 0.98, 0.99.$ We then evaluate the models 
chosen by cross validation on the $1072$ observations collected in the 12th period, 
of which $216$ are recorded as fraudulent. The results of this is summed up in Table
\ref{table:anls1} and Figure \ref{fig:anls1}, where the fraud loss is plotted as a 
function of $k/n$ for both the models, and both of the ways of setting the tuning 
parameter. For the penalised logistic regression models, it is evident from Figure 
\ref{fig:anls1} that the models chosen by the fraud loss are better than the model chosen by the AUC, at least up to around $k/n \approx 0.3,$ while for the boosted 
trees it is harder to see a clear difference between the two. The reported figures in 
Table \ref{table:anls1} show that the fraud loss performs better, both in terms of the 
average relative fraud loss aggregated over $k/n = 0.01, 0.02, \dots, 0.99$, and over 
the smaller selection of values $k/n = 0.16, 0.17, \dots, 0.25,$ for both of the models. 
The absolute figures reported in parentheses also show that penalised logistic 
regression in this case gave somewhat lower fraud loss than the boosted trees, 
which indicates that this model is perhaps a little better suited for the given setting.

\begin{table}
  \caption{Relative aggregated fraud loss, with aggregated fraud loss given in
  parentheses. Comparison of AUC and fraud loss. Dataset of VAT fraud from the
  Norwegian tax administration.
  \label{table:anls1}}
  \input{analys_dat1.tex}
\end{table}

%% file: analys_dat1.tex
\begin{tabular}{c|c|c}
  \hline
  Estimation model & Logistic & Additive trees\\  
  \hline
  \multicolumn{2}{c}{Average over all $k/n = 0.01, 0.02, \dots 0.99$:}\\
  \hline
  auc              &1.0412\,(0.6974) & 1.0328\,(0.6948)\\
  fraud            &1.0276\,(0.6911) & 1.0285\,(0.6930)\\
  \hline
  \multicolumn{2}{c}{Average over $k/n = 0.16, 0.17, \dots 0.24, 0.25$:}\\
  \hline
  auc              &1.0854\, (0.6361) &1.0301\,(0.6244)\\
  fraud            &1.0609\,(0.6218) &1.0269\,(0.6225)\\
  \hline
\end{tabular}

%% file: conclusion.tex
Statistical fraud detection consists in creating a system, that automatically selects
a subset of the cases that should be manually investigated. However, the 
investigator is often limited to controlling a restricted number $k$ of cases. In order
to allocate the resources in the most efficient manner, one should then try to select
the $k$ cases with the highest probability of being fraudulent. Prediction models that
are used for this purpose, must typically be regularised to avoid overfitting. In this 
paper, we propose a new loss function, the fraud loss, for selecting the complexity
of the prediction model via a tuning parameter. More specifically, we suggest an
approach where either a penalised logistic regression model, or an additive tree model 
is fitted by maximising the log-likelihood of a binary regression model with a 
logit-link, and the tuning parameter is set by minimising the fraud loss function. 

In a simulation study, we have investigated different ways of selecting the model
complexity with the fraud loss, taking the out-of-sample performance into account,
either by cross validation or bootstrapping. Based on this, repeated cross validation
with few folds seems to be the most favourable. In particular, we have opted for 
repeated 2-fold cross validation without stratification. Still, we recognise that 
stratification might be necessary if there are very few cases of fraud in the training 
data. 

Then, we carried out a larger simulation study, where we compared the performance 
of setting tuning parameters by cross validating the fraud loss, to cross validating the 
AUC. In these simulations, we saw that the fraud loss gave the best results in most 
cases, particularly when the proportion of the cases we to select is close to the 
marginal probability of fraud.

We have also illustrated our approach on a dataset of VAT fraud from the
Norwegian Tax Administration, making the same comparison as in the second
round of simulations. In this example, the fraud loss performed better than the AUC,
most substantially when fitting penalised logistic regression models, which were
also the model that were the most adequate for this application. 

We have focussed on two particular estimation methods and corresponding
definitions of the model complexity. The first is maximising the logistic
log-likelihood function, subject to ridge regularisation, where the penalty parameter
is the one to be chosen. The second is boosting for an additive tree model, where
the number of trees is the focus. The first could however be easily adapted to the
other types of regularisation, such as the lasso or the elastic net. For the second,
one might define the complexity in terms of for instance the size of each tree. 
One could also imagine using the fraud loss to select the complexity for other types
of binary classification models. Further, one might search for new divergences
to optimise that put more emphasis on estimating the higher probabilities
accurately, which is similar the the work of \citet{rudin2009p},
\citet{NIPS2012_4635} and \citet{eban2016scalable}. Another alternative is to adapt 
regression trees to the problem of picking a certain number of cases. This might
be done by fitting small trees directly combined with bagging, or by pruning a decision tree to minimise the fraud loss after growing the tree using a standard splitting 
criterion.

%% file: acknowledgements.tex
This work is funded by The Research Council of Norway centre Big Insight,
Project 237718. The authors would also like to thank Riccardo De Bin, for
his useful input, and participation in discussions. 